\documentclass[aps,12pt,final,notitlepage,oneside,onecolumn,nobibnotes,nofootinbib,superscriptaddress,noshowpacs,centertags]{revtex4}
\usepackage{amsmath,amssymb,epsfig,placeins,subfigure,wrapfig,indentfirst,makeidx}
\usepackage[T2A]{fontenc}
\usepackage[utf8x]{inputenc}
\usepackage{ucs}
\usepackage{amsmath}
\usepackage{mathtext}
\usepackage{amsfonts}
\usepackage{upgreek}
\usepackage[english]{babel}
\usepackage{amsfonts,amssymb}
\usepackage{color}
\usepackage{graphicx}
\usepackage{mathrsfs}
\usepackage{dsfont}

\begin{document}
\selectlanguage{english}

\title{Analysis of the Topology of the Kerr Metric}

\author{Alexander Shatskiy}
\affiliation{shatskiyalex@gmail.com}

\date{\today}

\begin{abstract}
The equations of motion of a test particle are integrated for the field of a rotating Kerr black hole (BH) (in accordance with~\cite{Carter1968}). 
Due to the lack of analytical transformations for the Carter–Penrose diagrams
(CPDs) for the Kerr metric, the topology of the Kerr BH is studied by analytical investigation of the equations of motion. 
Transformations for the CPDs for the Reisner–Nordström metric are analyzed. 
The problem of boundary conditions for the Reisner–Nordström topology is analyzed. 
A solution to this problem of boundary conditions is proposed. 
It is proved that, in the Reisner–Nordström topology, only one way to go to another universe is possible. 
For the Kerr topology, the possibility of the existence of an alternative transition 
to another universe that does not coincide with the universe for the ordinary transition is found. 
This alternative transition is performed through a surface with a zero radial coordinate (zero radius). 
Initial conditions for the falling particle are found that correspond to an alternative transition to another universe. 
The tidal forces acting on a falling body in the Kerr metric are estimated, and the possibility of the transition of the body
to other universes without being destroyed by tidal forces is proved.
\end{abstract}

\maketitle

\section{Introduction}
\label{introduction}

It has long been known that real black holes (BHs) that exist eternally (from the very origin of the 
Universe\footnote{We will write the word “universe” with a lowercase letter, meaning by this different parts of the big Universe, 
which combines different parts into the complex topology of the Universe.} 
are in a space with complex topology (see, for example, \cite{Frolov1998}). 
Henceforth, by real black holes we mean BHs that have either angular momentum (the Kerr solution), or electric charge 
(the Reisner–Nordström solution), or both (the Kerr–Newman solution).

\begin{figure}
\centering
\begin{minipage}[h]{0.39\linewidth}
\center{\includegraphics[width=0.99\linewidth]{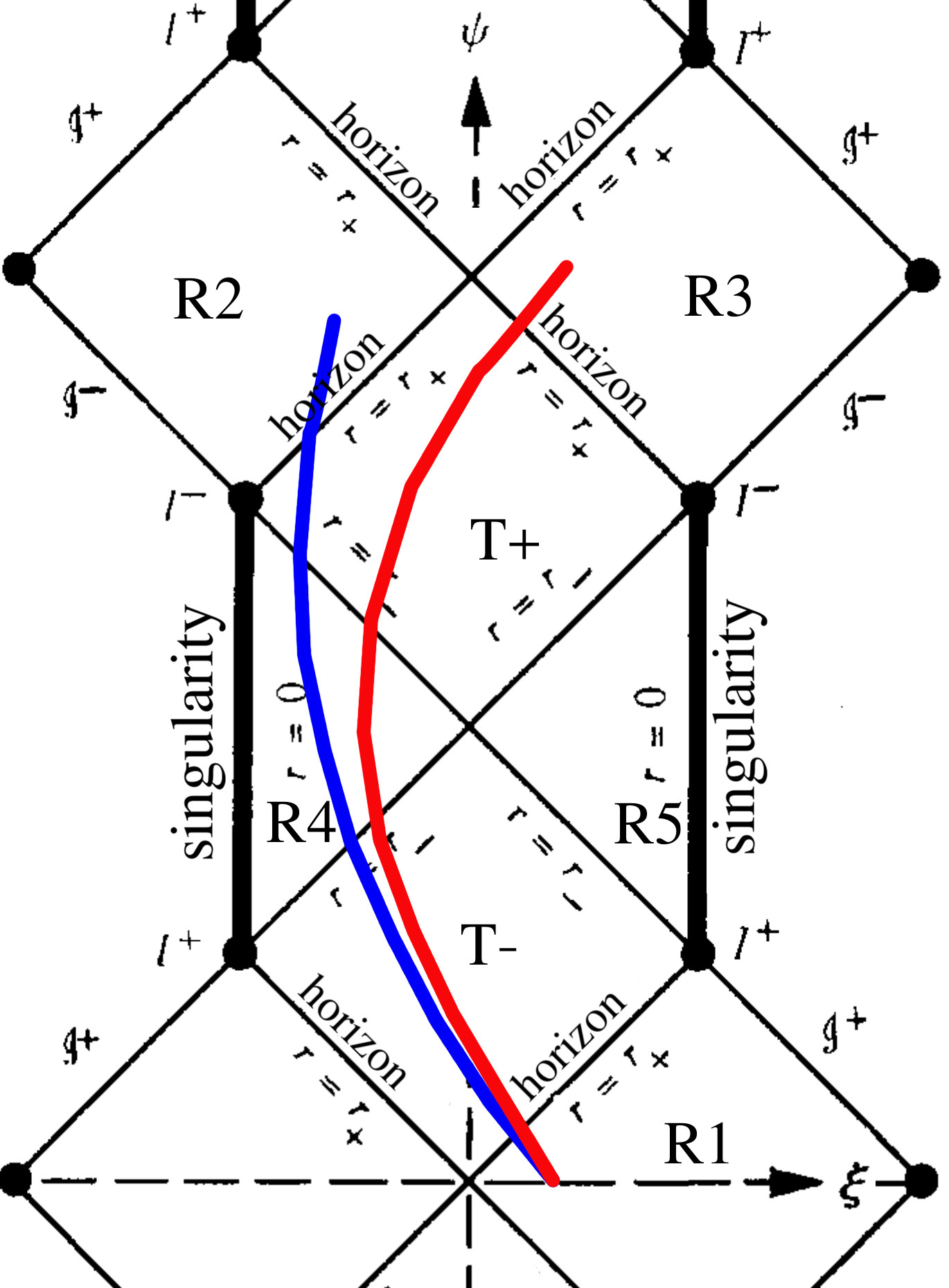} \\ a)}
\end{minipage}
\begin{minipage}[h]{0.29\linewidth}
\center{\includegraphics[width=0.99\linewidth]{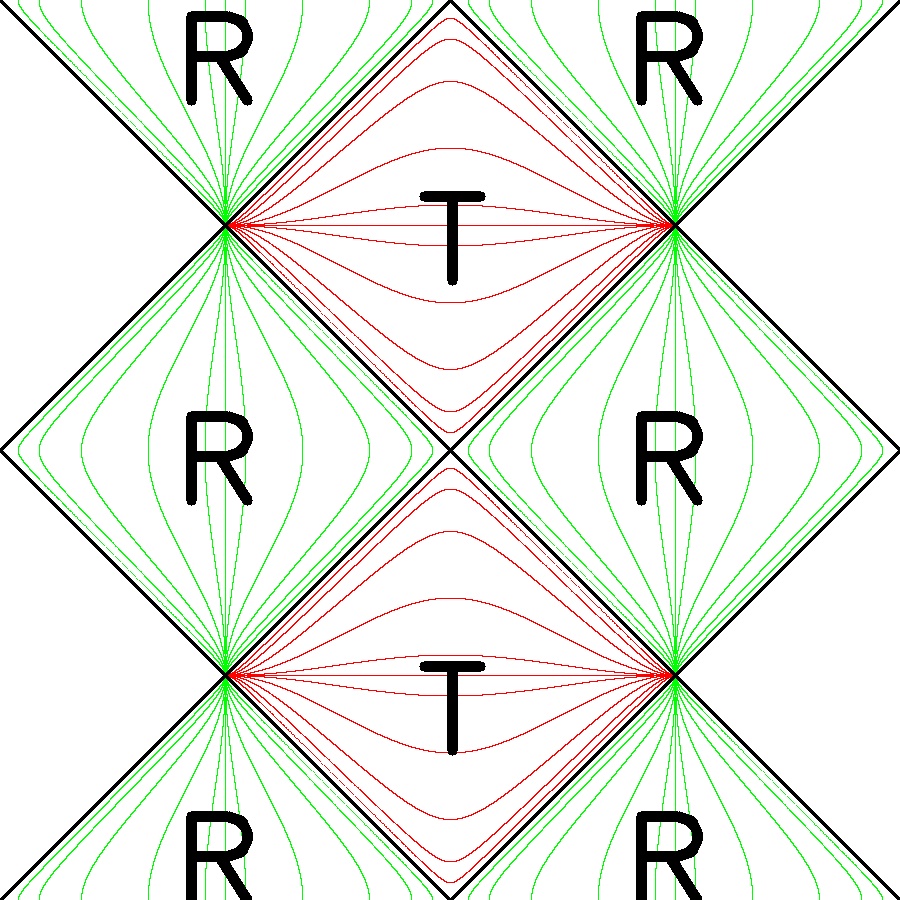} \\ b)}
\end{minipage}
\begin{minipage}[h]{0.29\linewidth}
\center{\includegraphics[width=0.99\linewidth]{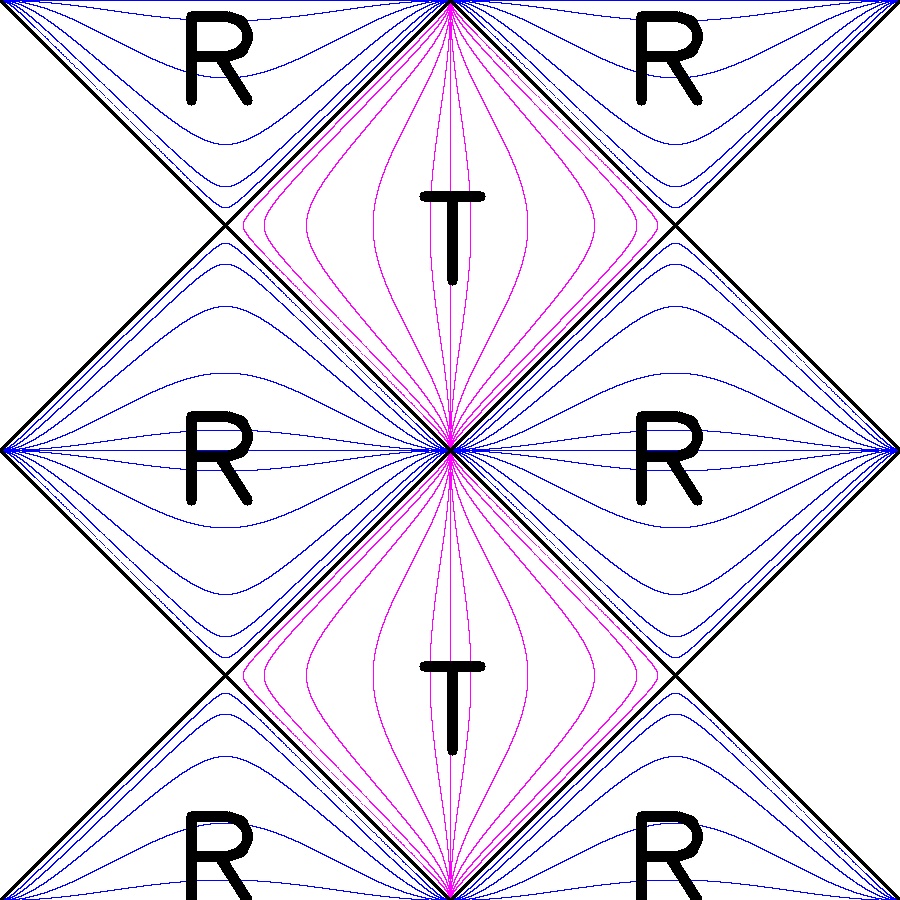} \\ c)}
\end{minipage}
\caption{{\\
(a) A Carter–Penrose diagram for a charged Reisner–Nordström BH. 
The singularity at ${r=0}$ is marked by bold vertical lines. 
Bold curves are possible world lines of particles. 
They enter the BH from region $R1$, then pass through the regions $T-$ , $R4$, and $T+$ and exit in regions $R2$ or $R3$ through different white holes into different universes. 
(b) Lines ${r=const}$ in $R$ and $T$ regions of the diagram. \\
(c) Lines ${t=const}$ in appropriate $R$ and $T$ regions of the diagram. \\ 
.\hrulefill
}}
\label{R1}
\end{figure}

For charged (but not rotating) BHs, there exists an analytical (and a graphical) Carter–Penrose diagram (CPD) (see Fig.~\ref{R1}), which shows that such BHs
exist in a space with complex topology that combines different universes (different parts of the big Universe).

In addition, charged BHs can be analytically studied completely; i.e., for example, one can construct 
trajectories of test particles in the gravitational field of these BHs. 
Moreover, these trajectories can be continued below the BH horizon and then further below the Cauchy horizon of this BH up to the turning point of the trajectory. 
In this case, a weak singularity on the Cauchy horizon does not prevent the continuation of 
these trajectories through the turning point (see below Fig.~\ref{R1}). 
After the turning point, the solution for the trajectory of a test particle moves to a new sheet (in another universe) and is smoothly stitched at the turning point. 

Here, the analytical indication of the turning point of the trajectory is the vanishing of the denominator in the integrand of the quadrature of the trajectory (see below). 
Of course, this point coincides with the turning point on the CPD when constructing this trajectory on it.

Unfortunately, there are no analytical transformations for two-dimensional CPDs of rotating BHs. 
This fact greatly complicates the analytical study of the complex topology of such BHs. 
Nevertheless, there is every reason to assume that the topology of rotating BHs is also complex, unlike, for example, the topology 
of uncharged and nonrotating (Schwarzschild) BHs.

The grounds for such assumptions are several facts that are considered in the present paper.

\section{Features of complex topology}
\label{features}

The first indication that the BH topology is complex is the presence of an $R$-region inside the BH, i.e., the existence of the Cauchy horizon in the solution. 
It was repeatedly shown earlier (see, for example,~\cite{Shatskiy2015_2}) that the singularity arising on the Cauchy horizon is
weak according to Tipler’s classification~\cite{Tip77}, i.e., integrable\footnote{The question is the convergence of the integral ${\int\limits\sqrt{K}ds}$, 
where $K$ is the Kretschmann scalar and $s$ is the proper time of a freely falling particle~\cite{Shatskiy2015_2}.}. 
In contrast to the strong (according to Tipler) singularity (which also necessarily exists inside any black hole), material bodies do not have time to be 
destroyed by tidal forces when passing through a weak singularity, since the action time of a weak singularity on a flying body tends to zero.

The second indication that the BH topology is complex is the “repulsion” of test particles from a strong singularity. 
As for the particles that are not test particles but are so heavy that they themselves have an opposite effect on the gravitation of the BH, 
most likely they are also repelled by the strong singularity of the BH; however, there is no mathematical (or numerical) confirmation of this fact yet.  
Moreover, “repulsion” means the absence of real particle trajectories that would end at a strong singularity. 

For charged BHs, this was proved analytically (see, for example,~\cite{Shatskiy2014}).

\section{Features of the Kerr metric}
\label{kerr1}

The solution for a rotating BH (the Kerr metric) in coordinates ${(t, r, \theta, \varphi)}$ has the form
\begin{eqnarray}
ds^2 = \frac{\Delta}{\rho^2}\left[dt - a\sin^2\theta\, d\varphi\right]^2 - \frac{\sin^2\theta}{\rho^2}\left[(r^2+a^2)d\varphi - a\, dt\right]^2 -
\frac{\rho^2}{\Delta}\, dr^2 - \rho^2\, d\theta^2 \, ,\label{ds2_1}\\
f(r) := (1 - r_h^{-}/r)(1 - r_h^{+}/r) \, , \label{f_r} \\
\rho^2 := r^2 + a^2\cos^2\theta \, .\label{rho2}
\end{eqnarray}
Here ${\Delta := r^2 f}$; and $a$ is the angular momentum $L$ of the BH, expressed in units of its mass $M$ (${a:=L/M}$);
we use dimensionless units: ${G=1}$ (gravitational constant) and ${c=1}$ (velocity of light).

By the asymptotic behavior of metric (\ref{ds2_1}) at infinity, we obtain:
\begin{eqnarray}
r_h^{-}+r_h^{+} = 2M\, , \label{r_h0}\\
r_h^{-}\cdot r_h^{+} = a^2\, ,\label{r_h1}\\
r_h^{\pm} := M \pm \sqrt{M^2 - a^2}\, . \label{r_h} 
\end{eqnarray}
Hence we obtain:
\begin{eqnarray}
\Delta := r^2 f = r^2 - 2Mr + a^2 \label{Delta}
\end{eqnarray}
The maximum possible value of the parameter a can be equal to the mass $M$ of a BH; then the inner Cauchy horizon touches the outer horizon in such a degenerate BH. 
A BH simply cannot rotate faster than for ${a=M}$; otherwise centrifugal forces would not allow gravitation to form such a BH. 
In reality, even this limit (when ${a=M}$) is not reachable; the maximum a can be approximately equal to ${0.998M}$~\cite{Thorne1974}.

As is known, strong singularities in the Schwarzschild solution and in the Reisner–Nordström solution are concentrated at the same point, 
which is the center of symmetry of the solutions. 
As for the Kerr solution (a rotating BH), here a strong singularity is concentrated on a ring. 
The geometrical locus of the points of this singular ring corresponds to the radial coordinate $r$ of the solution, which turns out to be zero at the singularity. 
By a strong singularity is meant the becoming infinite of the Kretschmann scalar $K$.

When calculating $K$, it turns out that the singularity in the Kerr metric arises not simply at ${r=0}$ (as was the case for a nonrotating BH), 
but also the condition ${\theta = \pi/2}$ is required; i.e., the singularity lies at the equatorial plane of the geometry at ${\rho =0}$~\cite{Krechman}:
\begin{eqnarray}
K := R_{ijkl}R^{ijkl} = \frac{48 M^2(r^6 - 15 a^2 r^4 \cos^2\theta + 15 a^4 r^2 \cos^4\theta -  a^6 \cos^6\theta)}{(r^2 + a^2\cos^2\theta)^{6}}
\label{Krech}\end{eqnarray}
Here ${R_{ijkl}}$ is the Riemann tensor, which is calculated based on the metric components $g_{ik}$~(see (\ref{ds2_1})).

Near the ring singularity (for ${\theta = \pi/2}$ and ${r\to 0}$), according to (\ref{Krech}), we have the asymptotics ${K\to 48 M^2/\rho^{6}}$. 
Consequently, the maximum of tidal forces corresponds to the minimum of $\rho$ on the trajectory, i.e., to the ring singularity.

Moreover, according to the rules of Riemannian geometry, the circumference of such a singular ring turns out to be equal to ${2\pi a}$, 
since the circumference is calculated from (\ref{ds2_1}) by the formula
\begin{eqnarray}
O = \int\limits_0^{2\pi} d\varphi \sqrt{|g_{\varphi\varphi}|} = 2\pi\sin\theta\sqrt{\frac{(r^2+a^2)^2-\Delta a^2\sin^2\theta}{r^2+a^2\cos^2\theta}}
\label{O}\end{eqnarray}
Therefore, for ${r=0}$ we obtain
\begin{eqnarray}
O_{(r=0)} = 2\pi a \sin\theta
\label{O_0}\end{eqnarray}
i.e., for ${r=0}$, the circle turns out to be the usual circle on a sphere of radius $a$. 
But this is exclusively due to the specific feature of the chosen coordinates in metric (1).

In addition to the definition of the circumference, we can also define the surface at ${r=const}$. 
The area of this surface is calculated from (\ref{ds2_1}) as the integral:
\begin{eqnarray}
S = \int\limits_0^\pi d\theta \int\limits_0^{2\pi} d\varphi \sqrt{|g_{\theta\theta}g_{\varphi\varphi}|} = 2\pi \int\limits_0^\pi \sin\theta\sqrt{(r^2+a^2)^2-\Delta a^2\sin^2\theta} \, d\theta
\label{S}\end{eqnarray}
Therefore, for ${r=0}$ we obtain 
\begin{eqnarray}
S_{(r=0)} = 2\pi a^2
\label{S_0}\end{eqnarray}
The quantity ${S_0 := 2\pi a^2}$ corresponds to the area of a two-sided circular membrane of radius $a$; 
i.e., in the definition of the surface, the quantity a behaves like the radius of a flat membrane. 
The difference in the definition of the radius at ${r=0}$ for the circle $0$ and for the surface area $S$ is due to the strong curvature of spacetime near the ring singularity.

\section{Proof of the fact that a test particle cannot reach a ring singularit}
\label{ring_rep}

The motion of a test particle in the field of a rotating BH is described by four integrals of motion (see~\cite{Thorn1977}, \S 33.5): 
the integrals for the mass $\mu$ of the particle, for its energy $\varepsilon$, for the projection $l_z$ of its angular momentum on the axis of rotation of the BH,
and the fourth (Carter) integral of motion $\Omega$. 
The equations obtained in this section will be used in Section~\ref{ring2} to plot the trajectory of a freely falling particle.

To simplify the calculations, it is convenient to write the equations in terms of the specific integrals of motion (divided by $\mu$): 
${\epsilon := \varepsilon/\mu}$ (the specific energy of the particle), ${h:=l_z/\mu}$ (impact parameter on the rotation axis of the BH), and ${\omega := \Omega/\mu^2}$  
(the specific Carter integral of motion).

Let us introduce the notation necessary for writing the equations:
\begin{eqnarray}
\Theta := \omega - \cos^2\theta\left[a^2(1-\epsilon^2) + h^2/\sin^2\theta)\right] \label{Theta}\, ,\\
P := \epsilon (r^2 + a^2) - ha \label{P}\, ,\\
\Re := P^2 - \Delta\left[r^2 + (h - a\epsilon)^2 + \omega\right]\, . \label{Re}
\end{eqnarray}
Let us write the equations of motion of test particles in the Kerr metric:
\begin{eqnarray}
\rho^2\frac{dt}{d\lambda} = a(h - a\epsilon\sin^2\theta) + (r^2+a^2)P/\Delta\, ,\label{dt} \\
\rho^2\frac{dr}{d\lambda} = \pm\sqrt{\Re} \, ,\label{dr}\\
\rho^2\frac{d\theta}{d\lambda} = \pm\sqrt{\Theta} \, ,\label{dtheta}\\
\rho^2\frac{d\varphi}{d\lambda} = (h - a\epsilon\sin^2\theta)/\sin^2\theta + aP/\Delta\, .\label{dvarphi}
\end{eqnarray}
Here, $\lambda$ is the affine parameter of the particle trajectory, the signs “$\pm$” in front of ${\sqrt{\Re}}$ and ${\sqrt{\Theta}}$ are chosen at the end with regard to the direction of motion of the particle. 
Further, to simplify the expressions, we remove the signs in front of ${\sqrt{\Re}}$ and ${\sqrt{\Theta}}$, implying the existence of these signs.

When determining the azimuthal coordinate $\varphi$ according to metric (\ref{ds2_1}), a singularity arises on the BH horizon. 
Because of this coordinate singularity, sometimes there is a misconception that any test particle in the Kerr metric makes an infinite number of revolutions 
around the BH before crossing the horizon. 
In fact, it is not a particle falling freely into a rotating BH that makes an infinite number of revolutions, 
but photons emitted by the particle to infinity from the horizon. 
The particle itself has time to make a rotation only through a finite angle before crossing the BH horizon. 
This contradiction is due to the fact that, in order to “escape” from the BH horizon to infinity, a photon needs infinite time. 
During this infinite time, gravimagnetic forces (that involve the locally inertial reference frame in rotation) have time to rotate the 
photon around the BH through an infinite angle. 
To eliminate these time and azimuthal coordinate singularities, we introduce renormalized time $T$ and azimuthal $\Phi$ coordinates 
(the Boyer–Lindquist coordinates), according to the definitions:  
\begin{eqnarray}
\frac{dT}{d\lambda} := \frac{r^2+a^2}{\rho^2}\left(\frac{dt}{d\lambda} - a\sin^2\theta\frac{d\varphi}{d\lambda}\right)  = \frac{(r^2+a^2) P}{\rho^2 \Delta}\, ,\label{dT}\\
\frac{d\Phi}{d\lambda} := \frac{r^2+a^2}{\rho^2}\left(\frac{d\varphi}{d\lambda} - \frac{a}{r^2+a^2}\cdot\frac{dt}{d\lambda}\right)  = \frac{h - a\epsilon\sin^2\theta}{\rho^2\sin^2\theta}\, .\label{dPhi}
\end{eqnarray}
Then, taking into account 
\begin{eqnarray}
\frac{d\lambda}{\rho^2} = \frac{dr}{\sqrt{\Re}} = \frac{d\theta}{\sqrt{\Theta}}\, ,\nonumber
\label{dlambda}\end{eqnarray}
we can write the integrals for calculating the particle trajectory as 
\begin{eqnarray}
\int\limits^\theta \frac{d\theta}{\sqrt{\Theta}} = \int\limits^r \frac{dr}{\sqrt{\Re}} \, ,\label{theta_r}\\
T = \int\limits^r \frac{(r^2+a^2) P\, dr}{\Delta\sqrt{\Re}} \, ,\label{T}\\
\Phi = \int\limits^\theta \frac{(h - a\epsilon\sin^2\theta)\, d\theta}{\sin^2\theta\sqrt{\Theta}} \, .\label{Phi}
\end{eqnarray}
Equation (\ref{theta_r}) is equivalent to Eq. (33.37а) from the book~\cite{Thorn1977}, and Eqs. (\ref{T}) and (\ref{Phi}) are combinations of Eqs. (33.37c) and (33.37d) 
according to the definitions (\ref{dT}) and (\ref{dPhi}).

Equations (\ref{theta_r}-\ref{Phi}) show that the condition under which a particle reaches a ring singularity with coordinates 
${r=0,\,\,\theta =\pi/2}$ is the nonnegativity of $\Re$ and $\Theta$ in the neighborhood of the singularity. 
According to expressions (\ref{Delta}, \ref{P}, \ref{Re}), we obtain
\begin{eqnarray}
\Re_{(r=0)} = -a^2\omega \label{sing1}
\end{eqnarray}
Let us prove that a test particle cannot reach a ring singularity by contradiction. 
First, consider the case of ${\Re >0}$ in the neighborhood of the singularity. 
In this case, according to (\ref{sing1}), $\omega$ should be negative; but then we find from (\ref{Theta}) that 
${\Theta_{(\theta = \pi/2)}=\omega<0}$ on the ring singularity, which contradicts Eqs. (\ref{theta_r}) and (\ref{Phi}).

Consider separately the case of ${\Re_{(r=0)}=0}$. 
In this case, ${\omega=0}$, and, according to (\ref{Theta}), we obtain ${\Theta_{(\theta = \pi/2)}=0}$. 
This degenerate case corresponds to trajectories only in the equatorial plane, since the integrals with respect to $\theta$ 
diverge logarithmically in the limit as ${\theta\to\pi/2}$. 
This means that, for ${\Re_{(r=0)}=0}$, the condition ${\theta = \pi/2}$, or ${\Theta = 0}$, must be satisfied for the entire trajectory.

The mathematical measure on the set of possible initial conditions for such trajectories (only in the equatorial plane and only for ${\omega=0}$) is zero on the set 
of all possible initial conditions. 
This situation is similar to that in the case of trajectories for test particles in the field of a charged Reisner–Nordström BH. 
In this case, only radial photons can reach the central singularity, and the measure of the initial conditions for 
radial photons is also zero on the set of all possible initial conditions (including nonradial photons). 

Thus, we have proved that test particles in the Kerr metric do not reach a strong ring singularity; i.e., a strong singularity as if “repels” particles.

\section{A Carter-Penrose diagram for a Reisner-Nordstr\"{o}m BH}
\label{KP_diagramm}

As already mentioned, due to the lack of central symmetry in the solution for a rotating BH, it is impossible to construct a two-dimensional analytical 
CPD for the Kerr solution\footnote{However, for any metric tensor (including the Kerr metric), it is possible to construct a graphical representation of a CPD 
(see, for example, Fig.~\ref{R3}). 
Such a graphical representation of a CPD is not always supported by the analytical transformations from this section, which are found only for centrally symmetric solutions.}. 
However, it is obvious from the graphical CPD in Fig.~\ref{R1}a that there can exist trajectories for a Reisner–Nordstr\"{o}m BH that pass from the same universe to different universes. 
It is not clear whether or not all such trajectories are physical; 
if yes, it is not clear to what initial conditions for the trajectories the boundary of transition to different universes corresponds. 
These questions will be addressed in the present section. 

Let us write the metric of a Reisner–Nordstr\"{o}m BH: 
\begin{eqnarray}
ds^2 = f(r)\,dt^2 - \frac{dr^2}{f(r)} - r^2\left( d\theta^2 + \sin^2\theta\, d\varphi^2 \right) \label{rn1} \\
r_h^{\pm} := M \pm \sqrt{M^2 - Q^2} \label{rn2}
\end{eqnarray} 
Here the function ${f(r)}$ is also defined by formula (\ref{f_r}); 
$M$ and $Q$ are, respectively, the mass and the charge of the BH; 
and $r_h^{\pm}$ are the radii of the horizons. 

Now, let us write a transformation from the coordinates ${(t, r)}$ to the “turtle” coordinates ${(\tau , \ell)}$, which correspond to a Reisner–Nordstr\"{o}m BH:
\begin{eqnarray}
f\,dt^2 - \frac{dr^2}{f} = f g^2\left(d\tau^2 - d\ell^2\right) \, , \quad g(r) := \pm\frac{dt}{d\tau}\, , \quad d\ell := \frac{dr}{|f|\cdot |g|} \label{dkp2}
\end{eqnarray}
Here we took into account that always ${dt>0}$ and ${d\tau >0}$, and ${sign(d\ell) = sign(dr)}$ by definition. 
According to~\cite{Carter1966}, we obtain
\begin{eqnarray}
g := sign(f) \label{g_r} = \pm 1
\end{eqnarray}
Now, we write an equation for a trajectory in a Reisner–Nordstr\"{o}m BH~\cite{Shatskiy2014}:
\begin{eqnarray}
U^r := \frac{dr}{ds} = \mp \epsilon\sqrt{1 - f (\epsilon^{-2} + h^2/r^2)} \label{rn4} \, ,\\
\frac{dr}{dt} = \mp |f| \sqrt{1 - f (\epsilon^{-2} + h^2/r^2)} \label{rn3}
\end{eqnarray}
The sign in (\ref{rn3}) is chosen according to ${sign(dr)}$ in the radial component of the four-velocity ${U^r}$: 
“minus” in the ${R1}$ and ${T^{-}}$ regions and “plus” in the ${T^{+}}$, $R2$ and $R3$ regions. 
In the region $R4$, ${U^r}$ vanishes and then changes its sign; at this moment, the sign in expression (\ref{rn3}) should also be changed. 
This point (the turning point) corresponds to the expression for the turning radius ${r_{turn}}$:
\begin{eqnarray}
\sqrt{1 - f_{(r_{turn})} (\epsilon^{-2} + h^2/r_{turn}^2)} = 0 \label{turn}
\end{eqnarray}
From (\ref{dkp2}) and (\ref{rn3}) we obtain
\begin{eqnarray}
d\ell = \frac{dr}{|f|} \label{d_ell} \\
\quad d\tau = dt =  \frac{\mp dr}{|f|\sqrt{1 - f (\epsilon^{-2} + h^2/r^2)}} \label{d_tau}
\end{eqnarray}
The transformations for the CPD from the “turtle” coordinates ${(\tau , \ell)}$ to the CPD coordinates 
${(\psi , \xi)}$ in accordance with~\cite{Carter1966} have the form 
\begin{eqnarray}
\tau + \ell = ctg\left(\frac{\psi + \xi}{2}\right) \, ,\quad \tau - \ell = tg\left(\frac{\psi - \xi}{2}\right) \label{dkp4} \\
f \left(d\tau^2 - d\ell^2\right) = F(\psi , \xi)\left( d\psi^2 - d\xi^2 \right) \label{dkp5}
\end{eqnarray} 
Denote ${u^{\pm} := 0.5\cdot(\psi\pm\xi)}$. 
According to the CPD in Fig.~\ref{R1}a, we can see that the coordinate${u^{+}}$ corresponds to the right-upward direction in the squares and the 
coordinate ${u^{-}}$ corresponds to the left-upward direction. 
When a particle moves along a trajectory on a CPD in the $T^{-}$ region, the coordinate u – runs through the values\footnote{The 
quantities ${u^{+}}$ and ${u^{-}}$ are defined up to ${\pi n}$, where $n$ is an integer, since the period of the functions  ${tg}$ and ${ctg}$ is $\pi$.} 
from ${-\pi/2}$ to ${+\pi/2}$, and the coordinate ${u^{+}}$ runs through the values from a constant ${c_1}$ to a constant ${c_2}$. 

When moving toward the turning point in the $T^{-}$ region, the range of variation of the coordinates ${\tau}$ and ${\ell}$ corresponds to the range of variation of the 
difference of coordinates ${(\tau - \ell)}$ -- from minus to plus infinity, and of the sum ${(\tau + \ell)}$ -- within a finite interval. 
This results from the fact that, when moving in this region, the integral of the combination ${(d\tau - d\ell)}$ tends to infinity if the limits of integration 
(with respect to the coordinate $r$) approach the horizons. 
At the same time, the integral of ${(d\tau + d\ell)}$ turns out to be finite even if the limits of integration lie on the horizons.

When moving after the turning point in the $T^{+}$ region, everything turns out to be opposite: 
the integral of ${(d\tau + d\ell)}$ tends to infinity if the limits of integration (with respect to the coordinate $r$) approach the horizons, 
while the integral of ${(d\tau - d\ell)}$ turns out to be finite even if the limits of integration lie on the horizons.

Accordingly, ${u^{+}}$ varies in the ranges\footnote{Here ${\{0 < c_1 < c_2 < \pi\}}$ and ${\{\pi/2 < c_3 < c_4 < 3\pi/2\}}$.} 
${[c_1,\, c_2]}$ for the $T^{-}$ region, ${[c_2,\, \pi]}$ for the $R4$ region, and ${[\pi,\, 2\pi]}$ for the $T^{+}$ region. 

The quantity ${u^{-}}$ varies in the ranges ${[-\pi/2,\, \pi/2]}$ for the $T^{-}$ region, ${[\pi/2,\, c_3]}$ for the $R4$ region, and ${[c_3,\, c_4]}$ for the $T^{+}$ region.

These ranges define the boundary conditions for the transformation (\ref{dkp4}). 
Figure~\ref{R1} shows that, in a CPD of a Reisner–Nordström BH, the trajectory of a freely falling body passes through the $T$ and $R$ regions, 
which are squares in the graphical representation of the CPD. 
It becomes clear from the boundary conditions for transformations (\ref{dkp4}) that, in the squares of the $T$ regions, the trajectory can pass only through the 
opposite sides of these squares and cannot pass through their adjacent sides. 
Therefore, for the Reisner–Nordstr\"{o}m BH, there can be only one path (trajectory in a CPD) to another universe. 
This unique path corresponds to the line leading to the universe ${R3}$ in the CPD in Fig.~\ref{R1}a, and a path to the universe ${R2}$ from the universe ${R1}$ is impossible.

Below we will prove that, for a Kerr BH in the Kerr solution for a rotating BH, there can exist another (alternative) path to another universe on a graphical CPD.

\section{Tidal forces on a trajectory}
\label{ring2}

\begin{figure}
\centering
\begin{minipage}[h]{0.49\linewidth}
\center{\includegraphics[width=0.99\linewidth]{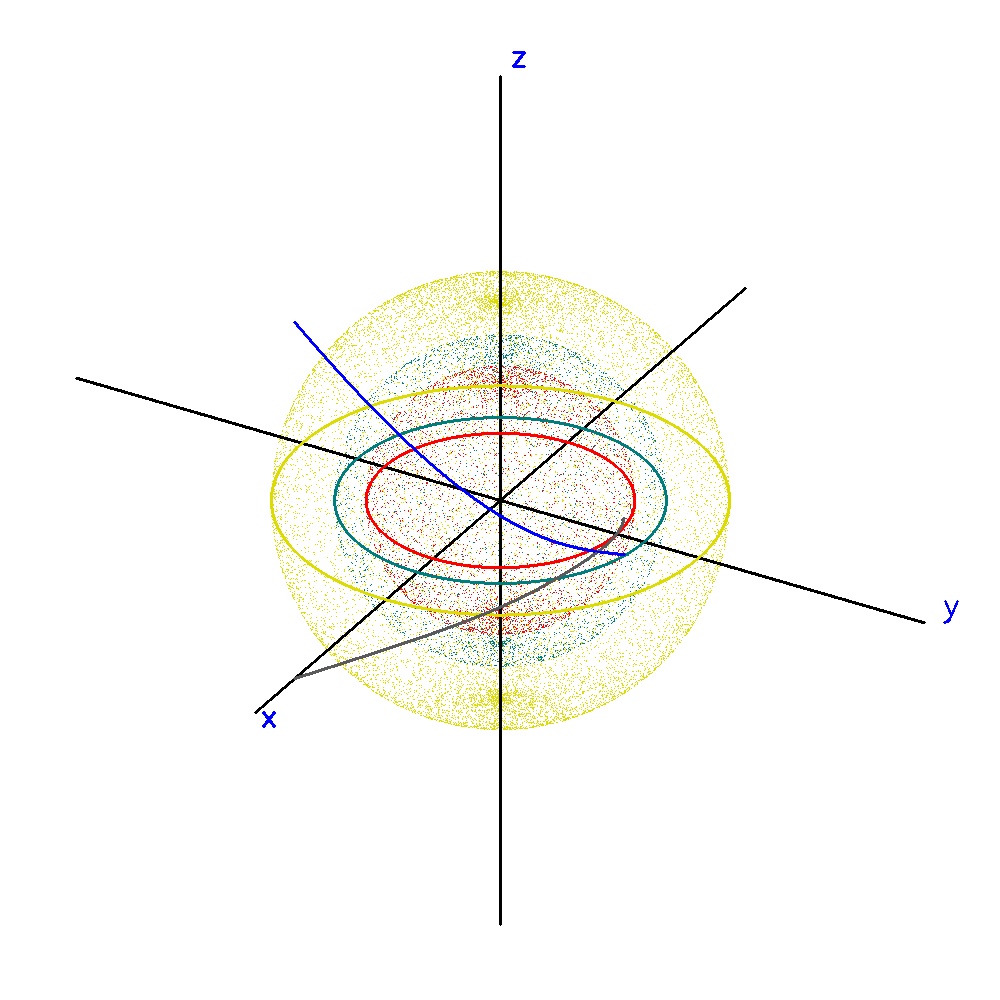} \\ a)}
\end{minipage}
\begin{minipage}[h]{0.49\linewidth}
\center{\includegraphics[width=0.99\linewidth]{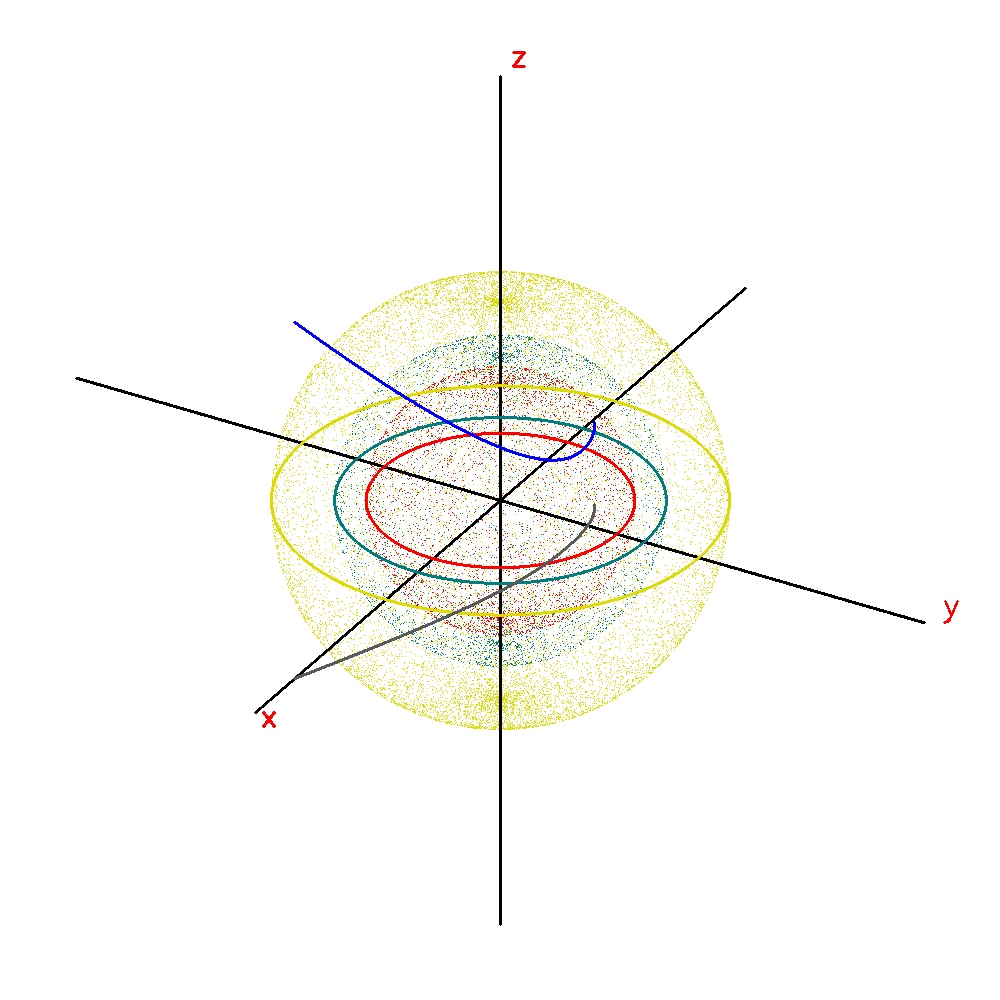} \\ b)}
\end{minipage}
\caption{{\\
Schematic view of the trajectory of a falling particle in a Kerr BH in Cartesian coordinates. 
The coordinates ${(x, y, z)}$ are defined in a standard way in terms of the spherical coordinates 
${(r_{eff}, \theta, \Phi)}$ (see~(\ref{theta_r}, \ref{Phi} and \ref{r_eff})). 
The red (inner) sphere with an area of ${4\pi r_{eff}^2 = 4\pi a^2}$ corresponds to the coordinate ${r=0}$; 
on this sphere, there is a ring singularity (red circle); 
the blue (intermediate) sphere with an area of ${4\pi [a^2+(r_h^{-})^2]}$ corresponds to the inner Cauchy horizon; 
the yellow (outer) sphere with area ${4\pi [a^2+(r_h^{+})^2]}$ corresponds to the outer horizon of the BH (see~(\ref{S} and \ref{r_eff})). 
The blue curve is the particle trajectory, and the gray curve is the projection of the trajectory onto the equatorial plane.\\ 
The initial conditions for the trajectories are as follows: 
${a=0.95M}$, ${\epsilon=2}$, ${h=-0.5M}$, ${\theta_0=\pi/4}$.\\
${\omega = 2M^2}$, ${r_{turn}>0}$ for variant (a); and ${\omega = -M^2}$, ${r_{turn}<0}$ for variant (b); the trajectory is constructed to the surface ${r=0}$.\\
.\hrulefill
}}
\label{R2}
\end{figure}

To graphically represent the trajectory of a falling body (Fig.~\ref{R2}), it is convenient to introduce the concept of effective radius ${r_{eff}}$:
\begin{eqnarray}
r_{eff} = \sqrt{r^2 + a^2}, 
\label{r_eff}\end{eqnarray}
the quantity ${4\pi r_{eff}^2}$ coincides with the surface areas calculated by formula (\ref{S}) on both horizons.

Despite the fact that the ring singularity itself is unattainable for particles falling into a rotating BH, 
the surface ${r=0}$ is attainable for particles with certain initial conditions. 
This surface is a set of turning points ${(r_{turn}=0)}$ for the trajectories that reach it. 
Similar to formula (\ref{turn}) for a Reisner–Nordstr\"{o}m BH, we can write a formula to determine a turning point for a Kerr BH:
\begin{eqnarray}
\Re (r_{turn}) = 0
\label{Kerr_turn}\end{eqnarray}
Here, the function ${\Re (r)}$ is defined by expression (\ref{Re}). 
After the turning point, the trajectory goes to a new solution sheet (into another universe).

For ${\omega = 0}$, according to (\ref{sing1}), ${r_{turn}=0}$; 
therefore, in the linear approximation with respect to small quantities ${|\omega| << M^2}$ and ${|r_{turn}| << |M|}$, we can write 
\begin{eqnarray}
r_{turn} \approx -\Re_{(r=0)} : \left.\frac{d\Re}{dr}\right|_{r=0} = \frac{a^2 \omega}{2M  (h-a\epsilon)^2}
\label{r_turn2}\end{eqnarray}
It is known that tidal forces in a BH are proportional to ${\sqrt{K}}$ (see~(\ref{Krech}) and \cite{Thorne2012}). 

Since the trajectory does not reach a singularity, the tidal forces acting on a falling body always remain finite. 
Thus, for a trajectory sufficiently remote from a ring singularity, the body can enter another universe through a blackwhite hole without destruction. 
This will depend on the mass of the BH and on the initial conditions of the trajectory.

The passage of a particle through a rotating black-white hole is possible either with (the turning point ${r_{turn}<0}$ (Fig.~\ref{R2}b)) or without 
(the turning point ${r_{turn}>0}$ (Fig.~\ref{R2}a)) reaching a zero membrane with radial coordinate ${r=0}$. 
A real turning point cannot be negative, since in the other universe, the particle satisfies new initial conditions.

\section{The problem of boundary condition for the Carter-Penrose diagrams}
\label{bound_cond}

The CPDs for a Reisner–Nordström BH show that the radius ${r_h^{-}}$ of the Cauchy inner horizon for the BH into which a particle falls should be the same as the 
radius of the Cauchy horizon of a white hole from which this particle then flies out into another universe (see~(\ref{r_h}). 
The radii ${r_h^{+}}$ of the outer horizons for black and white holes need not be the same, and the radii of the inner horizons should remain the same both 
before and after the passage of the particle. 
Since the particle has a mass (energy) ${\mu>0}$, the mass of the BH after the particle falls into it should increase, while the 
mass of the white hole should decrease after the particle leaves it. 
The radii of the horizons for black and white holes change accordingly.

When a particle with positive mass enters a BH, the radius ${r_h^{-}}$ of its Cauchy horizon always decreases; 
accordingly, when this particle flies out from a white hole, the radius ${r_h^{-}}$ of its inner horizon should also decrease (see~(\ref{r_h})): 
\begin{eqnarray}
\frac{dr_h^{-}}{dM} = 1 - \frac{M}{\sqrt{M^2-a^2}} - \frac{a^2}{M\sqrt{M^2-a^2}} < 0 \label{drc_dM}\, , \\
\frac{dr_h^{-}}{dL} = \frac{a}{M\sqrt{M^2-a^2}} \, . \label{drc_dL}
\end{eqnarray}
Formula (\ref{drc_dL}) takes into account the possible change in ${r_h^{-}}$ associated with a change in the angular momentum ${L=aM}$ of the BH. 
One can see that the change in ${r_h^{-}}$ associated with the passage of a particle is almost always nonzero. 
When a particle flies out into another universe, the change in ${r_h^{-}}$ for a white hole should be the same in magnitude and sign, which is very difficult 
to achieve by the choice of the initial conditions for the particle. 

Therefore, for positive masses, the conditions for the equality of the radii of the inner horizons before and after the passage of the particle are hardly compatible. 
This circumstance was the main obstacle to the development of the theory of black–white holes.

In Section 34.6~\cite{Thorn1977}, Misner, Thorne, and Wheeler considered three different possibilities for a particle falling into a real BH. 
The escape of this particle from a black–white hole was there the third possibility. 
However, for the above reason, this third possibility was rejected by the authors, just as the other possibilities (which were rejected for other reasons). 
Combinations of these three possibilities neither gave the desired result about the fate of a particle participating in the gravitational collapse.

However, if we assume that a particle in another universe flies out of a black–white hole with negative mass, then this would immediately solve the problem 
for the third possibility (the problem of equality of the radii of the inner horizons before and after the passage of the particle).

Indeed, after the escape of a particle of mass ${\mu>0}$ from a white hole with negative mass ${M_{wh}<0}$, the mass of the white hole decreases by ${\mu}$, 
i.e., ${\delta M_{wh}<0}$, but ${M_{wh}}$ increases in absolute value. 
This increase (in absolute value) will correspond to an increase in the radius of the Cauchy horizon by ${\delta r_h^{-}>0}$ for the white hole. 
Hence, this increase can easily be correlated with an increase by ${\delta r_h^{-}>0}$ for the BH into which the particle falls.

Henceforth, for brevity, we will call our universe positive, and the universe with negative masses 
(into which a particle with negative mass from our universe flies), negative. 
Since the magnitude of the radii of the BH horizons is determined in the general theory of relativity by the asymptotics of the metric at infinity, 
to eliminate the contradiction in the negative universe, it will suffice to put a minus sign in front of M everywhere in formulas 
(\ref{r_h0}), (\ref{r_h}), (\ref{Delta}) and (\ref{rn2}). 
For the same reason, the “less than” sign in formula (\ref{drc_dM}) remains valid for negative $M$.

We are still accustomed from school to the fact that the mass of any body should always be nonnegative. 
However, it is quite obvious that the sign in front of the mass in any equations can be chosen arbitrarily—this will not lead to the violation of physical 
laws and equations, provided that the signs in front of the masses in all other equations are also reversed. 
This situation is completely analogous to the situation with the signs of charges in electromagnetic theory: 
the reversal of the signs of all electric charges does not violate anything this is just a change of notation. 
A more detailed study of this topic was carried out in~\cite{Shatskiy2011}, where the authors showed that the motion of a particle with negative 
mass in our universe does not lead to any anomalies. 
In general, the sign and magnitude of mass in the universe is determined by the degree of inertia, i.e., by Mach’s principle. 
In support of Mach’s principle, Hawking wrote: “The observed isotropy of the microwave background indicates that the rotation of the Universe is very small, if any... 
This can be considered as an experimental confirmation of Mach’s principle”~\cite{Hawking1969}. 

In addition to solving the problem of matching boundary conditions for black–white holes, the use of negative masses gives additional advantages: 
gravitational repulsion by a white hole (with a negative mass) of a particle with positive mass that flies out from it 
can easily be explained, because bodies with masses of opposite sign repel each other.

A particle with positive mass cannot fall into a BH in a negative universe, 
since it will be gravitationally repelled by all heavy (negative) bodies in the negative universe. 
But if we somehow attach this positive particle to a body with negative mass in a negative universe, 
then this coupled pair can fall into a black–white hole in the negative universe and fly out in a positive universe\footnote{However, 
it is likely that any such connection will be destroyed by infinite gravitational forces inside the black–white hole.},  
similar to the fall of a positive particle into a BH in our universe. 

In this section, we made an assumption that the gravitational perturbations caused by the passage of a particle through a BH and the 
emission of gravitational waves associated with this passage can be neglected. 
This assumption is justified by the fact that the relative change in the radii of all horizons during the passage of a test particle is on the order of ${\mu/M}$  
and the relative change in these radii associated with the emission of gravitational waves by a test particle is on the order ${10^{-2}\mu^2/M^2}$, see~\cite{Martel2008}.

As is known, massive bodies move in space along trajectories, being attracted to the minima of the gravitational potential. 
In the main approximation of the two-body problem, these trajectories have the shape of Keplerian orbits (ellipses, hyperbolas, and parabolas). 
For bodies with negative mass in a similar gravitational field, nothing is essentially changed; 
the bodies move along Keplerian orbits, but now around the maxima, rather than the minima, of the gravitational potential. 
These maxima of the gravitational potential in deep Space correspond to the centers of intergalactic voids, and the distances between voids are usually hundreds 
of megaparsec. Particles with negative mass can accumulate in these voids. 
A similar situation should occur for particles with positive masses in a negative universe. 
The accumulation of particles with negative mass in voids can lead to cosmological effects similar to the effect of dark energy in cosmology.

Negative particles may accumulate near the Lagrange points in solar-type stellar systems, but most likely their velocities will be on the order of or greater 
than the average galactic velocity, about 300~km/s. 
Therefore, it is likely that negative particles will fly out without delay from parent galaxies and can be trapped only by the “gravitational hills” of the voids. 
A more detailed and quantitative analysis of these issues requires separate consideration.

\section{Alternative variants}
\label{ring3}

\begin{figure}
\centering
\begin{minipage}[h]{0.5\linewidth}
\center{\includegraphics[width=0.99\linewidth]{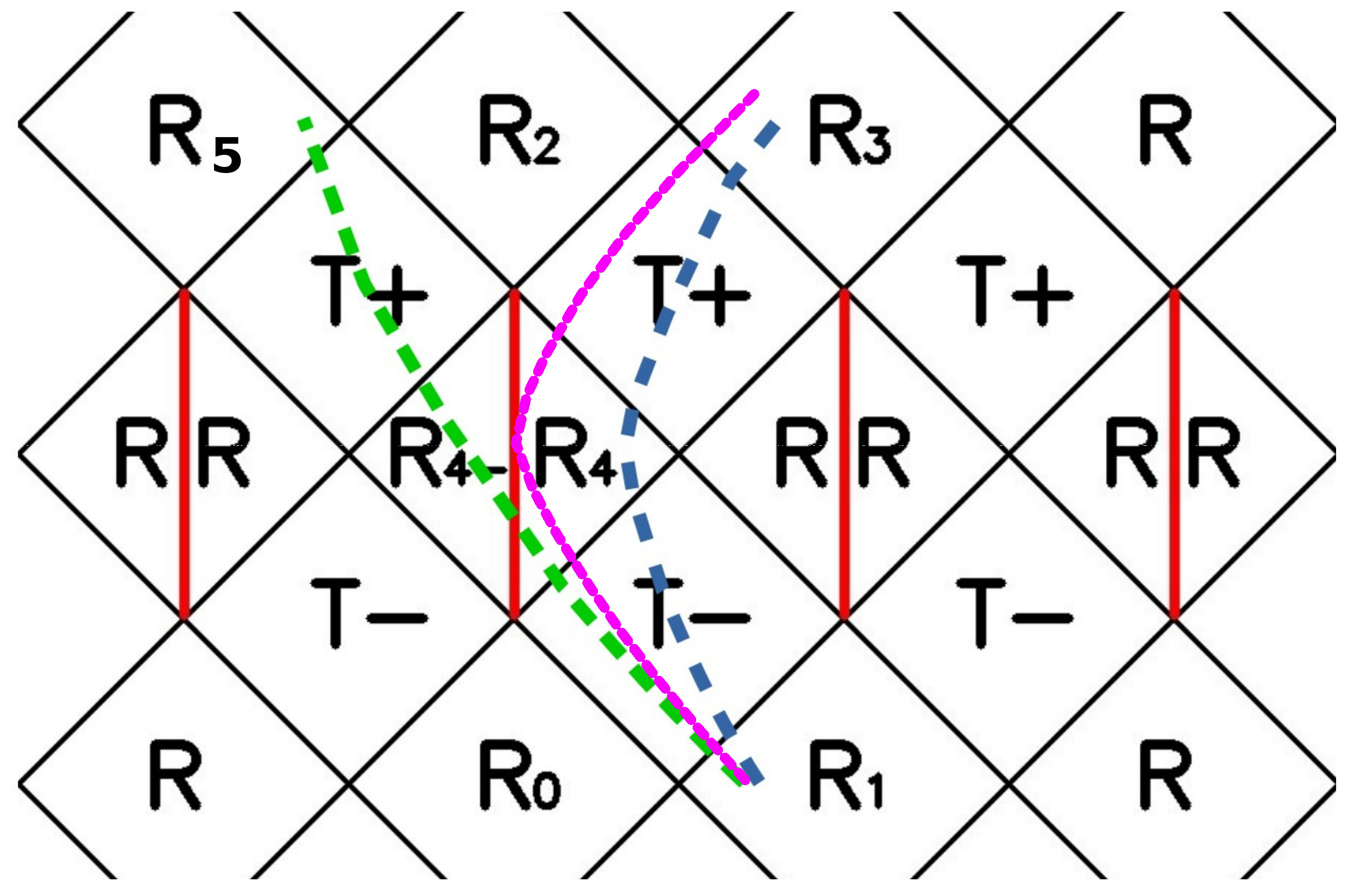}}
\end{minipage}
\caption{{\\
Carter–Penrose-type diagram for a Kerr BH with an alternative transition to another universe. 
The vertical bold lines mark the surfaces ${r=0}$. 
The standard transition to another universe is through the regions ${R_1 - T^{-} - R_4 - T^{+} - R_3}$  
(in accordance with the CPD in Fig.~\ref{R1} for a Reisner–Nordstr\"{o}m BH), 
the blue and violet dashed lines, and the alternative transition is performed through the regions ${R_1 - T^{-} - R_4 - R_{4-} - T^{+} - R_5}$, 
the green dashed line.\\
.\hrulefill
}}
\label{R3}
\end{figure}

The Kerr solution for a rotating BH, in contrast to the Reisner–Nordstr\"{o}m solution for a charged BH, 
provides another unique opportunity for moving to another universe. 
This possibility is associated with the fact that the surface area corresponding to the zero radial coordinate in the Kerr solution (\ref{ds2_1}) is greater than zero: 
${S_0 = 2\pi a^2}$ (see~(\ref{S})), as well as with the fact that a body can stay on this surface (membrane) without being destroyed by tidal forces, 
provided that this body is not at a singularity. 
Therefore, solution (\ref{ds2_1}) can pass through zero for the radial coordinate r and go further in the same direction provided that the radial 
component of the four-velocity ${U^r := \frac{dr}{ds}}$ is negative for a body falling into a BH at ${r=0}$ (Fig.~\ref{R3}).

It can be proved\footnote{This follows upon substituting ${dt}$, ${dr}$, ${d\theta}$ and ${d\varphi}$ from (\ref{dt}--\ref{dvarphi}), 
expressed in terms of ${d\lambda}$, into (\ref{ds2_1}).} 
that ${ds^2 = d\lambda^2}$. 
Then, according to (\ref{dr} and \ref{dtheta}), we have 
\begin{eqnarray}
U^r = \frac{dr}{d\lambda} = \mp\frac{\sqrt{\Re}}{\rho^2} \, ,\quad U^\theta = \frac{d\theta}{d\lambda} = \pm\frac{\sqrt{\Theta}}{\rho^2}
\label{dlambda_ds} \end{eqnarray}
In addition, according to (\ref{r_turn2}), the radius ${r_{turn}}$ can be negative only for ${\omega<0}$. 

Note that it makes sense to construct a trajectory at most to the zero radius $r$, and the negativity of ${r_{turn}}$ only indicates that an alternative transition 
is possible to another universe for the given initial conditions for the incident particle. 
To construct an extension of this trajectory to another universe, one should construct a new trajectory in reverse time in another universe and 
smoothly stitch together the two parts of the trajectory at ${r=0}$.

This makes clearer the physical meaning of the specific Carter integral $\omega$. 
According to (\ref{Theta} and \ref{theta_r}), the trajectory cannot intersect the equatorial plane for ${\omega<0}$. 
Since ${U_{\theta} = \pm\sqrt{\Theta}}$, from (\ref{dlambda_ds}) with ${\omega>0}$ and ${\theta = \pi/2}$, we have ${U_{\theta (\theta = \pi/2)} = \pm\sqrt{\omega}}$; 
i.e., for positive $\omega$, the root of the specific Carter integral has the meaning of the projection, orthogonal to the $z$ axis, of the specific 
angular momentum of the particle at the time when it intersects the equatorial plane. 
It is clear from (\ref{Theta}) that this projection is not an integral of motion. 
In the field of a rotating BH, this projection is not preserved—only the projection of the total angular momentum of the particle onto the $z$ axis, ${\mu h}$, is preserved.

If the trajectory passes through ${r=0}$ (to another universe), then, to eliminate confusion with the signs of $r$, we can change the notation of the sign of the 
radial coordinate, ${r\to -r}$, and carry out further calculations in the other universe in the same way as in our universe. 
This alternative variant is also interesting in that it does not require smooth stitching of the boundary conditions on the horizon, which was required in 
Section~\ref{bound_cond}. 
This is associated with the fact that, in the other universe, there is also a rotating BH with a zero membrane (at ${r=0}$) the area of which needs not be 
equal to the area of the membrane in our universe. 
Therefore, we should have no difficulty in stitching on this membrane.

Note that solution (\ref{ds2_1}) depends linearly on $r$ only in one place (in the term with  ${2Mr}$ in $\Delta$); in other places, the dependence on $r$ is quadratic. 
Therefore, the other universe can be both positive and negative. 
In the negative case, there is no need in changing the notation of the sign of the radial coordinate, ${r\to -r}$, since solution (\ref{ds2_1}) retains its form.

Due to the continuity requirement for the metric, we can't arbitrarily “stitch“ two parts of spacetime together at any given place. 
This also applies for the "stitching“ on different sides of the BH membrane. 
The “stitching" can be performed by different methods. 
And with these different methods of "stitching“ there would be a certain variant of solution. 
Now it is not clear: what which of these variants of the solution was chosen by nature. 
It is most convenient to perform this research in the Kerr-Shield coordinates ${(x, y, z)}$ 
(see for example~\cite{Carter1968_2}, \cite{Kerr1964}, \cite{Hawking1973} \S 5.6). 

So, for the Kerr BH solution near the membrane, there are three variants:\\
1) Either the transition through the membrane corresponds to the smooth transition to the region ${R_{4-}}$ -- to negative values $r$ (see green dashed line on Fig.~\ref{R3}), 
at the same time the BH mass is still positive,  
and then, in this case, a new universe ${R_5}$ opens behind the membrane, in which necessarily contains closed world geodesic lines (variant 1).\\ 
2) Or, as in the first variant, the transition through the membrane corresponds to a smooth transition to the region ${R_{4 -}}$, to the negative $r$ values, 
but in this new universe ${R_5}$ the BH mass becomes negative, and there are no closed world geodesic lines (variant 2).\\ 
3) Otherwise, the transition through the membrane corresponds to the same internal region ${R_4}$;  
i.e., in fact, the reflection (see violet dashed line on Fig.~\ref{R3}) on the coordinate $r$,  
which now remains everywhere non-negative, there are no closed world geodesic lines, and the metric is broken on the membrane (smoothness violation),
and the membrane itself will be a material surface with a negative mass, and the modulo of this mass is the order of the entire BH mass (variant 3). 

For the first and for the second variants, the membrane is insubstantial, and for the third variant, the circumference around the membrane in the inner region,
and the passing through the membrane returns us to the same inner region ${R_4}$. 
In the third variant, there is no alternative passage to another universe,
since the particle as if reflected from the membrane and finally fly out into the universe $R_3$ as it was discussed in the previous sections. 

Personally, I prefer the second or the third variant, because, as it appears, the negative masses are not something more exotic,
than wormholes, or black-white holes. 
Moreover, the negative masses and phantom matter in theory do not violate any physical laws, 
and they can even explain the above-mentioned anti-gravity effect under the inner BH horizon, 
can explain some effects in cosmology – for example, the effect of dark energy (rapidly universe expanding), as well as the birth of the universe itself. 
Let me remind that the transition through a black-white hole to another universe is also possible without the reaching of the membrane by a falling body, 
but this will be the universe $R_3$, which does not correspond to the universe $R_5$ behind the black hole membrane 
(in which the body falling into while  passing through the membrane in the first or second variants). 

And here, in an obscure way, the beauty and complexity of non-trivial topology in General relativity for black-white holes begins to unfold...

\section{Discussion}
\label{discuss}

Meanwhile, the theory of gravity cannot give an unambiguous answer to the question of the possibility of existence of black–white holes in our Universe. 
Mathematically, only the “eternal” complete solution for black–white holes is known (it is depicted in the CPD in Fig.~\ref{R1}a); 
it is a solution that exists eternally, from the very origin of the Universe (from the moment of the Big Bang). 
It is not actually clear how the complete solution (corresponding to some Carter-Penrose diagram) should look like for a BH formed as 
a result of a star collapse (that is, after the Big Bang).

For now, it is clear that:\\  
-- any real BH has not only an outer horizon, but also the inner Cauchy horizon;\\ 
-- physical bodies are able to overcome a weak singularity on the Cauchy horizon without destruction;\\ 
-- a strong singularity inside a real BH will repel any matter below the inner horizon;\\ 
-- the tidal forces are finite, because the trajectory of a falling body does not touch a strong singularity;\\ 
-- this matter cannot escape back from the same BH any longer. 

However, it is not yet clear what will happen with this matter—a complete solution for the BHs born after the Big Bang is not yet found, nor self-consistent 
complete solutions are found that take into account the back reaction of the falling matter on the gravity of the whole system. 

We have obtained the following results:\\ 
-- We have proved that a falling test particle cannot reach a strong ring singularity inside a Kerr BH.\\ 
-- We have found a solution to the problem of boundary conditions for the Reisner-Nordstr\"{o}m topology.\\ 
-- We have proved that only one variant of transition to another universe is possible in the Reisner-Nordstr\"{o}m topology.\\ 
-- For the Kerr topology, we have found that there can exist an alternative transition to another universe that does not coincide with the universe 
for the ordinary transition.\\ 
-- We have found initial conditions for a falling particle that correspond to the alternative transition to another universe.\\ 
-- We have estimated the tidal forces acting on a falling body in the Kerr metric and proved the possibility of the passage of a body to other universes 
without being destroyed by tidal forces. 

We have also proposed a hypothesis about the existence of “negative universes” and negative masses, which allows one to stitch the boundary conditions in 
black and white holes. 

Despite the attractiveness of the arguments in favor of the hypothesis proposed on the existence of alternative transitions to another universe, a final 
mathematical proof of such transitions has not yet been found.

\hrulefill 

Translated by I. Nikitin
\end{document}